\documentclass[%
reprint,
superscriptaddress,
longbibliography,
 amsmath,amssymb,
prb,
]{revtex4-1}

\usepackage[dvipsnames]{xcolor}
\usepackage{graphicx}
\usepackage{dcolumn}
\usepackage{multirow}
\usepackage{bm}
\usepackage{bbold}
\usepackage[colorlinks=true,linkcolor=blue]{hyperref}
\usepackage{mathrsfs}
\usepackage{lipsum}
\usepackage{float}

\begin{document}


\title{Multiphase tin equation of state using density functional
  theory}

\author{Daniel A. Rehn}
\affiliation{Computational Physics Division, Los Alamos National Laboratory, Los Alamos, NM 87545, USA}
\affiliation{Theoretical Division, Los Alamos National Laboratory, Los Alamos, NM 87545, USA}
\affiliation{Center for Nonlinear Studies, Los Alamos National Laboratory, Los Alamos, NM 87545, USA}
\author{Carl W. Greeff}
\affiliation{Theoretical Division, Los Alamos National Laboratory, Los Alamos, NM 87545, USA}
\author{Leonid Burakovsky}
\affiliation{Theoretical Division, Los Alamos National Laboratory, Los Alamos, NM 87545, USA}
\author{Daniel G. Sheppard}
\affiliation{Computational Physics Division, Los Alamos National Laboratory, Los Alamos, NM 87545, USA}
\author{Scott D. Crockett}
\affiliation{Theoretical Division, Los Alamos National Laboratory, Los Alamos, NM 87545, USA}

\begin{abstract}
  We perform density functional theory (DFT) calculations of five
  solid phases and the liquid phase of tin. The calculations include
  cold curves of the five solid phases, phonon calculations in the
  quasi-harmonic approximation over a range of volumes for each solid
  phase, and DFT-based molecular dynamics (DFT-MD) calculations of the
  liquid phase.  Using the DFT results, we construct a tabular
  multiphase SESAME equation of state for tin, referred to as SESAME
  2162.  Comparisons to experimental data are made and show a high
  level of agreement in isobaric data, isothermal data, shock data,
  and phase boundary measurements, including measurements of the melt
  curve. The 2162 EOS will be useful for hydrodynamics simulations and
  has been designed with an eye toward hydrodynamics simulations that
  incorporate materials strength models and allow for modeling of the
  kinetics of phase transitions.
\end{abstract}

\maketitle

\section{Introduction\label{sec:intro}}
Tabular equations of state (EOS) are important for both a basic
understanding of materials properties and for hydrodynamics
applications.  As hydrodynamics codes are developed to incorporate
materials strength models~\cite{preston2003model,barton2011multiscale}
and to account for kinetic effects in simulating phase
transitions,\cite{greeff2016model,andrews1971calculation,wills2016implementing}
the need for an accurate underlying multiphase EOS becomes
increasingly important.  Here we focus on the development of a new
tabular multiphase
SESAME~\cite{lyon1992sesame,chisolm2005constructing} EOS for tin,
referred to as SESAME 2162. The EOS can be thought of as a successor
to the previous tin EOS, SESAME 2161.~\cite{greeff2005sesame} The new
EOS includes four solid phases and the liquid phase.

To generate SESAME 2162, we performed density functional theory (DFT)
calculations of the solid phases of tin and the liquid phase,
including static lattice energy (cold curve) calculations, phonon
calculations in the quasi-harmonic approximation, and DFT-based
molecular dynamics (DFT-MD) calculations of the liquid phase,
including calculations of the melt curve.  These calculations provide
data in regions of the state space where experimental data is
unavailable and thereby help to constrain construction of the final
EOS.

Tin is known to exist in at least 5 different solid phases, summarized
in Table~\ref{tab:1}. Throughout, we refer to the phases by their
corresponding Greek letter. Note that the $\alpha$, $\beta$, and
$\gamma$ naming conventions are all standard. Here we also use
$\delta$ to refer to the bcc phase, which has infrequently been
referred to as the $\sigma$ phase in the
past,~\cite{molodets2000thermodynamic} and we use $\epsilon$ to refer
to the hcp phase. The Greek letters are chosen in ascending order
according to the pressure at which each phase becomes stable.

Although we performed DFT calculations for all five solid phases, we
include only the $\beta$, $\gamma$, $\delta$, and $\epsilon$ phases in
the 2162 EOS, since the $\alpha$ phase is only stable below room
temperature and at pressures below 1
GPa~\cite{young1991phase,tonkov2018phase,nikolaev1973characteristics}
with limited accessibility in compressive and shock experiments, which
are the main focus of hydrodynamics codes that use the EOS. In
addition, the transition from $\beta\rightarrow\alpha$ is slow and
therefore unlikely to show up in experiments.~\cite{anderson2000phase,cornelius2017phenomenon}

\begin{table}[!ht]\centering
  \begin{ruledtabular}
  \begin{tabular}{c|c|r|c|c}\footnotesize
    phase & lattice type & space group & space group \# & \# atoms/cell \\ \hline
    $\alpha$ & diamond & $Fd\bar 3m$    & 227 & 2 \\\hline
    $\beta$  & bct     & $I4_1/amd$   & 141 & 4 \\\hline
    $\gamma$ & bct     & $I4/mmm$     & 139 & 2 \\\hline
    $\delta$ & bcc     & $Im\bar 3m$    & 229 & 1 \\\hline
    $\epsilon$ & hcp   & $P6_3/mmc$   & 194 & 2 \\
  \end{tabular}
  \caption{Summary of the crystal structures and space groups of the 5
    known solid phases of tin. Greek letters are used to refer to the
    phases in ascending order, from low to high pressure. DFT
    calculations are performed using a mixture of primitive and
    conventional unit cells, with the number of atoms per cell
    indicated in the last column.}
  \label{tab:1}
  \end{ruledtabular}
\end{table}

Many experiments have been performed on tin, including isobaric
measurements,~\cite{kammer1972elastic,touloukian1975thermal,swanson1953standard,alchagirov2000temperature,wang2003precise,hultgren1973selected,thewlis1954thermal,prasad1955study}
isothermal diamond anvil cell (DAC)
measurements,~\cite{olijnyk1984phase,liu1986compressions,desgreniers1989tin,plymate1988pressure,cavaleri1988pressure,salamat2011dense,salamat2013high,stager1962high,barnett1966x,vaidya1970compressibility}
shock
experiments,~\cite{walsh1957shock,al1958dynamic,mcqueen1960equation,al1962shock,marsh1980lasl,al1981shock,hu2008shock,rice1958compression}
dynamic compression experiments,~\cite{lazicki2015x} and measurements
of the solid-solid phase
boundaries.~\cite{kingon1980redetermination,xu2014phase,barnett1963x,hinton2020response} Measurements
of the phonon dispersions for the $\alpha$, $\beta$, and $\gamma$
phases have been also been
made.~\cite{price1971lattice,rowe1967crystal,ivanov1987lattice,ivanov1995fermi,giefers2007phonon}
In addition, the melt curve has been measured in a variety of
different ways, including shock-induced
melt~\cite{mabire2000shock,mabire2000shock2,seagle2013shock,briggs2019coordination,hereil2002temperature}
and resistive and/or laser heating in a DAC or compressive
piston.~\cite{dudley1960experimental,weir2012high,la2019high,jayaraman1963melting,narushima2007pressure,briggs2012melting,briggs2017high,schwager2010melting}
The available melt curve data show a large variability in the range of
measurements. More recently, studies on liquid spallation and
fragmentation have also been
performed.~\cite{de2007experimental,signor2010investigation,anderson2000phase}
In addition, a wide range of theoretical calculations of tin have been
performed, including DFT-based cold curve
calculations,~\cite{corkill1991theoretical,hafner1974ab,ihm1981equilibrium,cheong1991first,christensen1993body,christensen1993density,aguado2003first,yu2006ab,cui2008first,yao2011prediction}
phonon
calculations,~\cite{wehinger2014diffuse,rowe1965lattice,chen1967group,pavone1998alpha}
and molecular dynamics
calculations,~\cite{bernard2002first,zhang2011general,ravelo1997equilibrium,soulard2020observation}
and a variety of equations of state for tin have been proposed over
the
years.~\cite{greeff2005sesame,buy2006thermodynamically,cox2006multi,khishchenko2008equation,cox2015fitting}

The rest of the paper is organized as follows: in
Sec.~\ref{sec:opensesame}, we provide an overview of OpenSesame, the
program used to generate SESAME 2162.  In Sec.~\ref{sec:dft} we
provide details on the calculations of the cold curves and phonon
calculations of the solid phases, and of the DFT-MD calculations of
the liquid phase.  We also describe how these calculations are used to
determine an initial set of parameters for models used in OpenSesame
to construct the 2162 EOS. In Sec.~\ref{sec:results}, we discuss how
the parameters from DFT are modified (when necessary) to obtain
agreement with experimental data and create the 2162 EOS.  We show
comparisons of the 2162 EOS to a variety of experimental measurements,
including isobaric data, isothermal DAC data, shock data, solid-solid
phase boundary measurements, and measurements of the melt curve. While
we only summarize the main results here, a more comprehensive
description of the process used to incorporate DFT calculations into
materials models used in OpenSesame, as well as a more detailed
description of how model parameters are adjusted to fit to
experimental data, is provided in Ref.~\onlinecite{rehn2020using}.

\section{Overview of OpenSesame\label{sec:opensesame}}
The OpenSesame software is a useful tool for generating tabular
materials equations of state. The multiphase capability in
OpenSesame~\cite{chisolm2005constructing} relies on individual phase
EOS tables and evaluates the phase with the lowest Gibbs free energy
at a given temperature and pressure. Mixed phases are handled in a
self-consistent manner, as described below. The multiphase capability
allows for an arbitrary number of materials phases to be included in
the final EOS table.  For the individual phase tables, OpenSesame uses
a decomposition of the total Helmholtz free energy $F(V,T)$ for each
phase into three pieces,
\begin{equation}
  F(V,T) = F_\mathrm{cold}(V) + F_\mathrm{ion}(V,T) + F_\mathrm{el}(V,T),
  \label{eq:F}
\end{equation}
where $F_\mathrm{cold}$ is the energy associated with a cold curve
($T=0$ static lattice energy), $F_\mathrm{ion}$ is the free energy
associated with ionic motion, and $F_\mathrm{el}$ is the electronic
free energy. OpenSesame uses a variety of simple materials models to
inform on these three contributions to the free energy, making it
necessary to determine model parameters from the DFT calculations
and/or experimental data. In constructing the 2162 EOS, we first
determined model parameters entirely from DFT calculations, as
discussed in Sec.~\ref{sec:dft}.  Because the DFT calculations
resulted in an EOS that was not in perfect agreement with experimental
data, small modifications to the model parameters were made, as
described in Sec.~\ref{sec:results}.

Model parameters for the solid phases determined by DFT are outlined
in Sec.~\ref{sec:dft}.  In Sec.~\ref{sec:coldcurves}, we discuss cold
curve calculations that inform directly on $F_\mathrm{cold}$, which we
treat with a Vinet-Rose model.~\cite{vinet1987temperature} In
Sec.~\ref{sec:phonons}, we show how phonon calculations can be used to
determine parameters in the Debye model~\cite{debye}, which informs on
$F_\mathrm{ion}$ for the solid phases. The Debye model relies on the
Debye frequency $\nu_D$, or equivalently the Debye temperature
$\Theta_D = h\nu_D/k_B$. Also necessary for $F_\mathrm{ion}$ is a
Gr\"{u}neisen model, also discussed in Sec.~\ref{sec:phonons}, that
provides information about the volume dependence of the Debye
frequencies.  DFT-based phonon calculations allow for the
determination of all parameters relevant to the Debye and
Gr\"{u}neisen models for the solid phases. The remaining term
$F_\mathrm{el}$ is determined using the Thomas-Fermi-Dirac (TFD)
method~\cite{chisolm2003thomas,thomas1927calculation,fermi1927metodo,dirac1930note}
and does not require additional parameterization from DFT
calculations.

For the liquid phase, the same decomposition in Eq.~\ref{eq:F} is
used. The liquid free energy is defined with respect to that of a
reference solid phase, which we denote $F_{\rm rs}$. This may be one
of the actual solid phases, or may be hypothetical. We define a
scaling temperatature $T_{\rm sc}(V)$, which marks the transition from
solid-like to liquid-like behavior.  For $T \leq T_{\rm sc}$, the
liquid free energy $F_{\rm l}$ is related to that of the reference
solid by a constant entropy shift $\Delta S$ and a volume-dependent
energy shift $\Delta F_{\rm cold}(V)$,
\begin{equation}
F_{\rm l}(V,T) = F_{\rm rs}(V,T) - \Delta S T + \Delta F_{\rm cold}(V) ;
  \,\,\, T \le  T_{\rm sc}(V),
\label{f_liq_lowt}
\end{equation}
where $F_{\rm rs}$ is parameterized the same way as the other solid phases.
For $T > T_{\rm sc}(V)$ the excess specific heat is assumed to be a function
of the scaled temperature
\begin{equation}
c_V(V,T)/N k_B  = f(T/T_{\rm sc}(V)) +3/2 \, , 
\label{cv_scaled}
\end{equation}
with $f(1) = 3/2$ and $f(x) \rightarrow 0$ as $x \rightarrow \infty$.
Eqs. \ref{f_liq_lowt} and \ref{cv_scaled} allow one to determine the
free energy at all temperatures,
\begin{align}
F_{\rm l}(V,T) = &F_{\rm l}(V,T_{\rm sc}(V)) 
               - S_{\rm l}(V,T_{\rm sc}(V)) (T-T_{\rm sc}(V))\nonumber\\
               & - \int_{T_{\rm sc}}^T \mathrm{d}T_2
                \int_{T_{\rm sc}}^{T_2} \mathrm{d}T_1 \frac{c_V(V,T_1)}{T_1} ;
               \,\,\, T >  T_{\rm sc}(V)
\label{f_liq_hight}
\end{align}
The details of the $T$-dependence of $c_V$ are described in
Ref.~\onlinecite{chisolm2005extending}.  We define $T_{\rm m}(V) =
\Delta F_{\rm cold}(V)/\Delta S$.  Then the volume dependence of the
two functions $T_{\rm sc}(V)$ and $T_{\rm m}(V)$ is given by
\begin{eqnarray} 
-{\mathrm{d} \ln T_{\rm m}(V)\over \mathrm{d} \ln V} & = & 2 \Gamma_{\rm m}(V) - 2/3 \label{tm_vdep} \\
-{\mathrm{d} \ln T_{\rm sc}(V)\over \mathrm{d} \ln V} & = & 2 \Gamma_{\rm rs}(V) - 2/3 
\label{ts_vdep}
\end{eqnarray} 
where the Gr\"uneisen parameters $\Gamma_{\rm m}$ and $\Gamma_{\rm
  rs}$ have the same functional form as those of the solid phase Debye
temperatures, as given in Eq.~\ref{eq:Gamma}, discussed in the next
section. Note that the effective Gr\"uneisen parameter for $T_{\rm
  sc}$ as defined in Eq.~\ref{ts_vdep} must be the same as the Debye
Gr\"uneisen parameter for the reference solid, $\Gamma_{\rm rs} =
-\mathrm{d} \ln \theta_{\rm rs}/\mathrm{d} \ln V$, in order for the
pressure to reach the ideal gas limit at high
temperature.\cite{grover1971liquid}

The liquid free energy has, in addition to the standard solid phase
parameters, two additional parameters for the volume dependence of
$\Gamma_{\rm m}$, as well as $\Delta S$ and reference values for
$T_{\rm m}$ and $T_{\rm sc}$.  This somewhat complicated formulation
was adopted to facilitate creating liquid models that obey standard
Lindemann scaling \cite{grover1971liquid} in the case of a single
solid phase. For that case, the reference solid is the actual solid
phase, the various Gr\"uneisen parameters are the same, and the
reference values of $T_{\rm m}$ and $T_{\rm sc}$ are the same and are
equal to the melting temperature at the reference volume. The earlier
SESAME 2161 tin EOS was made with the liquid ``normal" in this sense
with respect to the high pressure $\gamma$ phase. For SESAME 2162,
there are more solid phases, we have more data on the high pressure
melting curve, as well as DFT-MD data on the liquid EOS, and thus were
not able to adopt this simplification. When developing the 2162 EOS,
we explored three options for determining a cold curve for the liquid,
as described in Ref.~\onlinecite{rehn2020using}.  Because melt is
possible from the $\beta$, $\gamma$, and $\delta$ phases of tin across
a wide compression range, we determined a cold curve that interpolates
across the three solid phases, also described in
Ref.~\onlinecite{rehn2020using}.


In the construction of SESAME 2162, we produce an equilibrium table
that includes mixed phase regions. The mixed phase regions are defined
by the following set of equations,
\begin{eqnarray}
P_a(V_a,T) & = & P_b(V_b,T) \nonumber \\
G_a(V_a,T) & = & G_b(V_b,T) \nonumber \\
\lambda_a V_a + \lambda_b V_b & = & V \nonumber \\
\lambda_a E_a(V_a,T) + \lambda_b E_b(V_b,T)  & = & E 
\label{coexist}
\end{eqnarray}
where $a$ and $b$ denote the two coexisting phases, $P$, $G$, $V$, and
$E$ are the pressure, specific Gibbs free energy, volume, and internal
energy, respectively, and $\lambda_a$ denotes the mass fraction of
phase $a$. For the great majority of states, there are no solutions to
the conditions (Eqs.~\ref{coexist}) satisfying $0 \le \lambda_i \le 1$
and $\sum_i \lambda_i =1$, and the state is a pure phase with the
lowest Helmholtz free energy at the given $V$ and $T$.  The algorithm
for constructing equilibrium tables is described in
Ref.~\onlinecite{chisolm2005constructing}.



The initial set of model parameters determined from DFT calculations
provide a useful starting point for the determination of the
OpenSesame model parameters described above. However, some adjustments
are needed to agree fully with experimental data. The adjustments made
are described in Sec.~\ref{sec:results}.

\section{DFT Calculations\label{sec:dft}}
\subsection{Cold curves\label{sec:coldcurves}}
All DFT calculations were computed using the Vienna \emph{ab-initio}
Simulation Package
(VASP),~\cite{kresse1993ab,kresse1994ab,kresse1996efficiency,kresse1996efficient}
version 5.4.4. All calculations use the projector augmented wave (PAW)
method~\cite{blochl1994projector,kresse1999ultrasoft} using either 14
valence states, 4d$^{10}$5s$^2$5p$^2$ (cold curves and phonons) or 4
valence states, 5s$^2$5p$^2$ (DFT-MD of the liquid).  The cold curves
were computed by relaxing the structures with multiple restarts to
ensure convergence of the atomic positions and lattice constants. The
relaxations were performed using the Methfessel-Paxton
scheme~\cite{PhysRevB.40.3616} with a 400 eV plane-wave energy cutoff.
Subsequently, total energy calculations were performed using fixed
ions and lattice constants with tetrahedral integration of the
Brillouin zone and a 520 eV plane-wave energy cutoff. We also computed
cold curves of tin using the all-electron code
RSPt~\cite{wills2010full} with different relativistic
treatments~\cite{rehn2020dirac,rehn2020relativistic} and found the
results to be in close agreement with the VASP results, indicating the
adequacy of using PAW method for treatment of the core states.

\begin{figure}
  \includegraphics[width=0.45\textwidth]{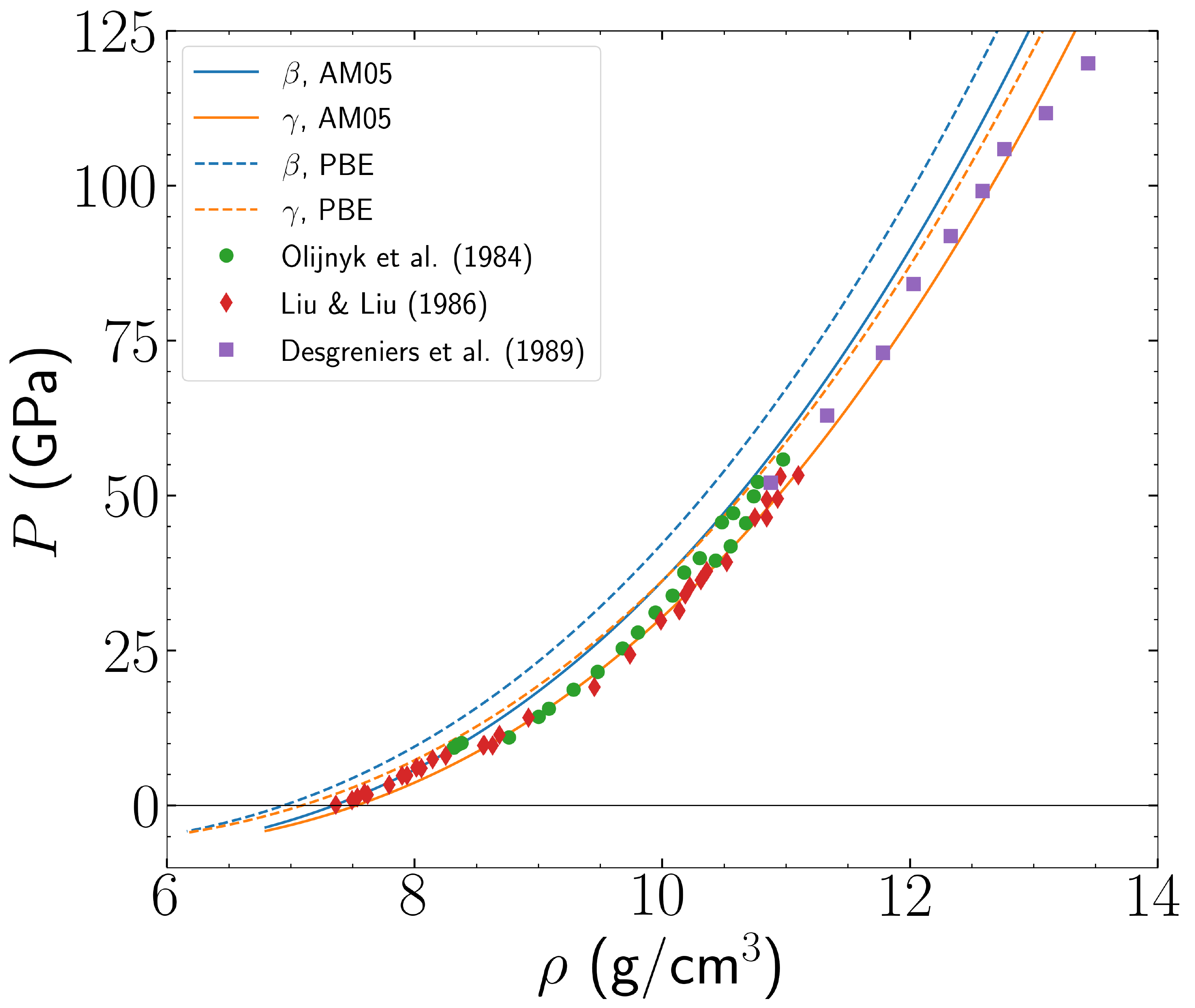}
  \caption{Cold curves for the $\beta$ (blue) and $\gamma$ (orange)
    phases of tin using the AM05 (solid lines) and PBE (dashed lines)
    exchange-correlation functionals. Experimental DAC data are from
    Refs.~\onlinecite{olijnyk1984phase,liu1986compressions,desgreniers1989tin}.}
  \label{fig:1}
\end{figure}

Cold curves were computed using the AM05~\cite{am05} and
PBE~\cite{pbe} exchange-correlation functionals. As shown in
Fig.~\ref{fig:1}, the AM05 results for the $\beta$ and $\gamma$ phases
are in very close agreement with experimental room temperature DAC
measurements from
Refs.~\onlinecite{olijnyk1984phase,liu1986compressions,desgreniers1989tin}
(note that the phase transition from $\beta$ to $\gamma$ occurs at
approximately 8.5 g/cm$^3$), whereas the equilibrium density $\rho_0$
predicted by PBE is substantially lower.  For this reason, we chose to
use the AM05 functional to determine the initial set of 2162 model
parameters, which include the AM05 cold curves, AM05 phonon
calculations, and DFT-MD calculations using AM05. The only exception
to this are DFT-MD calculations of the melt curve using the $Z$
method~\cite{belonoshko2006melting,burakovsky2014z,burakovsky2015ab},
which were computed separately using the PBE
functional.~\cite{Leonid-private-communication}.  We present the melt
curves only in $P$-$T$ space (see Fig.~\ref{fig:4}). These results
should be expected to be quite similar to what would be obtained using
the AM05 functional, since the primary difference between AM05 and PBE
is an offset in the density, while the pressure as a function of the
compression $\rho/\rho_0$ for the two functionals are quite similar,
as can be seen in Fig.~\ref{fig:1} and as verified by highly similar
bulk moduli computed from the cold curves (see
Ref.~\onlinecite{rehn2020using} for details).

The Vinet-Rose cold curve model was used to fit to the AM05 data of
each phase. The results for $\rho_0,E_0,B$, and $B'= dB/dP$ are
provided in Table~\ref{tab:2}.  Note that for cold curve calculations
of the $\beta$, $\gamma$, and $\epsilon$ phases, the lattice constants
are allowed to relax at each volume to allow for a volume-dependent
$c/a$ ratio.  In the case of the $\beta$ and $\epsilon$ phases, $c/a$
remains relatively constant over the full compression ranged studied,
varying by only a few percent across a compression range of
approximately 0.9 to 3.0.  On the other hand, the $\gamma$ phase shows
a large variation in the $c/a$ ratio, starting at $c/a=0.88$ at
$\rho_0=7.53$ g/cm$^3$ and converging to $c/a=1$ above 9.38 g/cm$^3$,
thereby indicating that the bcc $\delta$ phase has the lower energy
above a compression ratio of approximately 1.25. The $\gamma$-$\delta$
phase transition is observed to occur between 10.5--11 g/cm$^3$
(compression of approximately 1.4--1.45) in recent isothermal DAC
measurements,~\cite{salamat2013high} which is in agreement with the
compression range at which the enthalpies of the $\gamma$ and $\delta$
phases computed using the AM05 cold curves cross. Additional details
regarding cold curve comparisons and the enthalpy are provided in
Ref.~\onlinecite{rehn2020using}, and comparisons of the room
temperature 2162 isotherms to experimental DAC data are provided in
Sec.~\ref{sec:results}.

We also point out that at non-zero temperatures, the electronic free
energy $F_\mathrm{el}$ can be determined through the use of Fermi
smearing in DFT calculations. This is analogous to cold curve
calculations, but with the Fermi smearing width set according to the
temperature $k_BT$. Although this can provide useful information, we
did not see any appreciable differences in the resulting phase
boundaries when including Fermi smearing to treat the electronic free
energy at a range of temperatures and volumes. In OpenSesame, the
electronic free energy is treated with the TFD
model,~\cite{chisolm2003thomas,thomas1927calculation,fermi1927metodo,dirac1930note}
which works over a wide range of densities and temperatures. Similar
to the use of Fermi smearing, the TFD method results in only small
quantitative differences in the resulting solid-solid phase
boundaries, thereby indicating that the TFD method can safely be used
for the solid phases of tin.

\subsection{Phonons\label{sec:phonons}}
\begin{table*}[!ht]\centering
  \begin{ruledtabular}
  \begin{tabular}{c|l|l|l|l|l|l|l}\footnotesize
    phase & $\rho_0$ (g/cm$^3$) & $E_0$ (J/g) & $B$ (GPa)  & $B'$ & $\Theta_D$ (K)& $\Gamma_\mathrm{ref}$& $\Gamma'_\mathrm{ref}$\\\hline
    $\beta$    & 7.336 (7.4375) & 0           & 52.74 (58.0) & 5.369 & 161.0         & 1.98 (2.42) & -1.31 \\ \hline
    $\gamma$   & 7.525          & 6.04 (36.0) & 52.32        & 5.344 & 121.7         & 2.48 (2.45) & -1.82 (-2.20) \\ \hline
    $\delta$   & 7.561          & 12.55 (54.0)& 52.64        & 5.383 & 114.2 (117.5) & 2.54 (2.51) & -1.73 (-2.10) \\ \hline
    $\epsilon$ & 7.531          & 3.05 (44.1) & 52.68        & 5.291 & 121.1 (124.6) & 2.51 (2.48) & -1.85 (-2.24)\\ \hline
    liquid     & (7.470)        & (24.0)      & (52.7)       & (5.540) & (131.0)         & (2.45) & (-3.50) \\
  \end{tabular}
  \caption{Model parameters used to construct the 2162 EOS. The
    DFT-calculated values using the AM05 functional are displayed
    first. If modifications were made to the parameter to better agree
    with experimental data, the final value used in the 2162 EOS is
    listed in parentheses next to the DFT-calculated value. Note that
    $\Theta_D$, $\Gamma_\mathrm{ref}$, and $\Gamma'_\mathrm{ref}$ for
    each phase are calculated at a reference density chosen to be
    $\rho_\mathrm{ref} = 7.581$ g/cm$^3$ for all phases. All liquid
    parameters are listed in parentheses because there is not a unique
    way to determine the values strictly from DFT calculations. The
    values chosen are ones that were found to best agree
    simultaneously with the DFT-MD data in Fig.~\ref{fig:4} and
    Hugoniot data in the liquid phase in Fig.~\ref{fig:8}. Also note
    that we use the liquid-specific reference parameters $T_{\rm m} =
    T_{\rm sc} = 580$ K, $\Gamma_{\rm m} = \Gamma_{\rm rs} = 2.45$,
    and $\Delta S = 0.9$ kJ/g/K.}
  \label{tab:2}
  \end{ruledtabular}
\end{table*}

\begin{figure}
  \includegraphics[width=0.45\textwidth]{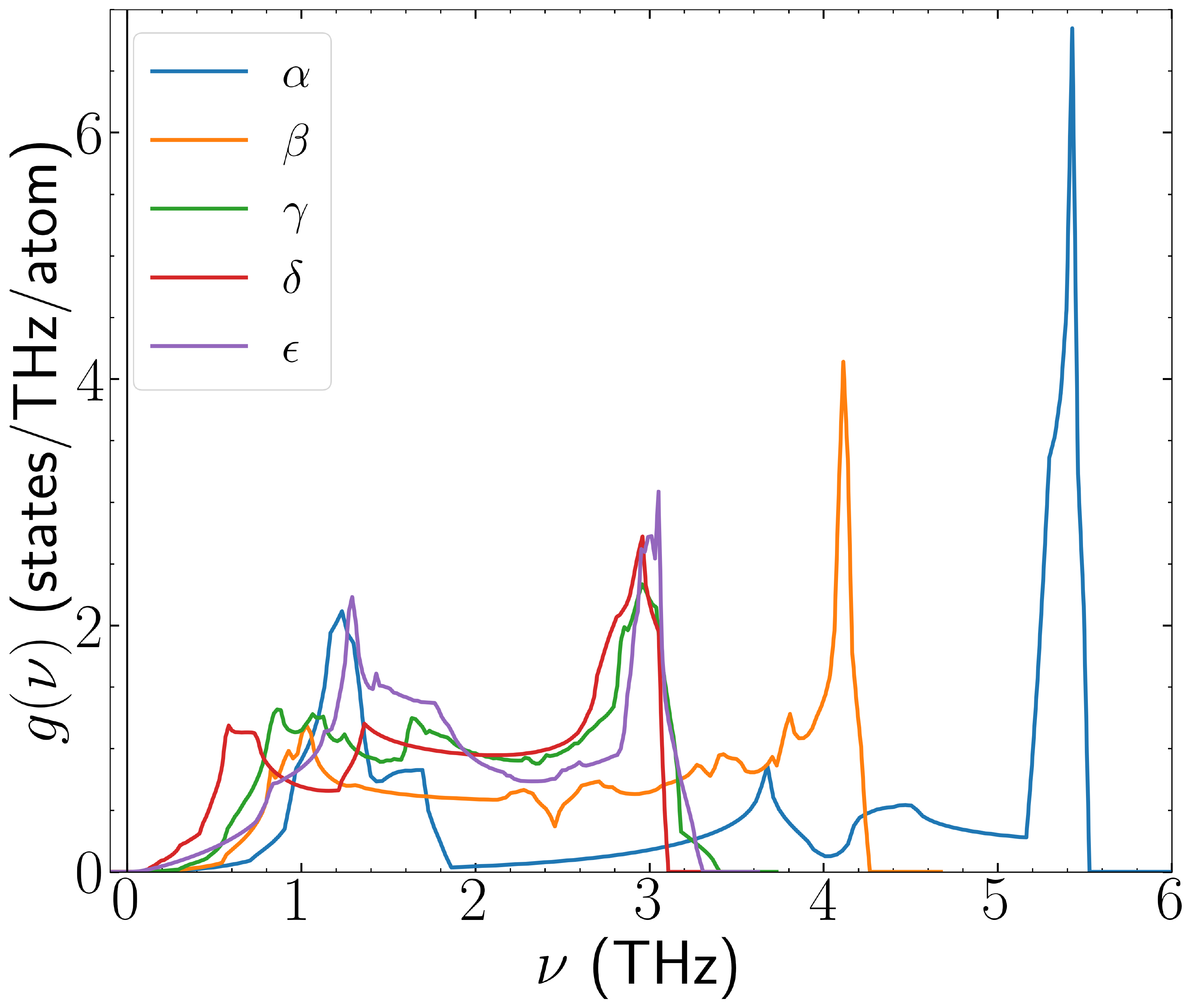}
  \caption{Phonon densities of states $g(\nu)$ for the different
    phases of tin at their respective equilibrium volumes using the
    AM05 functional.}
  \label{fig:2}
\end{figure}

Phonon calculations were performed in the quasi-harmonic
approximation~\cite{dove1993introduction} using a frozen phonon
supercell approach in VASP.  The Phonopy~\cite{phonopy} package was
used to generate supercells and atomic displacements for each solid
phase. The plane-wave energy cutoff, k-mesh size, and supercell size
were tested at the AM05-predicted equilibrium volume for each phase
individually to ensure structural stability of each phase and to
ensure a converged phonon density of states. We use a minimum 350 eV
plane-wave energy cutoff for each solid phase and used the method of
Methfessel and Paxton~\cite{PhysRevB.40.3616} to sample the Brillouin
zone. In all calculations we determine the k-mesh of the Brillouin
zone using a $\Gamma$-centered
Monkhorst-Pack~\cite{monkhorst1976special} scheme. The supercells were
constructed using the primitive or conventional cells listed in
Table~\ref{tab:1}, including a 54 atom cell with a $6\times6\times6$
k-mesh ($\alpha$ phase), a 64 atom cell with an $8\times8\times6$
k-mesh ($\beta$ phase), a 54 atom cell with an $8\times8\times8$
k-mesh ($\gamma$ phase), a 64 atom cell with a $6\times6\times6$
k-mesh ($\delta$ phase), and a 54 atom cell with a $10\times10\times7$
k-mesh ($\epsilon$ phase).  At volumes below the equilibrium volume,
we use the same k-mesh size to ensure that results remain converged
with respect to k-mesh size.  The phonon densities of states $g(\nu)$
for each phase at their respective equilibrium volumes are shown in
Fig.~\ref{fig:2}.

The phonon calculations provide a rigorous way to determine model
parameters in OpenSesame used to construct the full tabular EOS. We
use the Debye model~\cite{debye} and the `generalized CHART D'
Gr\"{u}neisen model~\cite{thompson1974improvements} within OpenSesame
to construct the 2162 EOS. Typically moments of $g(\nu)$ are used to
determine the Debye frequencies of the solid
phases.~\cite{chisolm2003test,chisolm2005extending} For tin, we found
that the zeroth moment typically gave free energies $F(T)$ that
matched the phonon free energy well over a wide range of temperatures
and volumes, whereas the free energy predicted using first and second
moments to determine the Debye frequencies did not match the phonon
$F(T)$ results very well. At the same time, evaluation of the zeroth
moment can be prone to numerical errors when a very small and
otherwise negligible density of states is present around zero
frequency (see Ref.~\onlinecite{rehn2020using} for details). This
motivated us to develop an alternative method for determining the
Debye frequencies for each phase and each volume based on a
minimization scheme, motivated by matching the Debye free energy to
the phonon free energy as closely as possible.  Given the phonon or
Debye density of states $g(\nu)$, the free energy at fixed volume $V$
is
\begin{equation}
  \small
  F(T; V) = \int_0^{\infty} g(\nu)\bigg({1\over 2}h\nu+ k_B
  T\ln\big[1-e^{-h\nu/k_BT}\big] \bigg)\,\mathrm{d}\nu.
\end{equation}

\begin{figure}
  \includegraphics[width=0.45\textwidth]{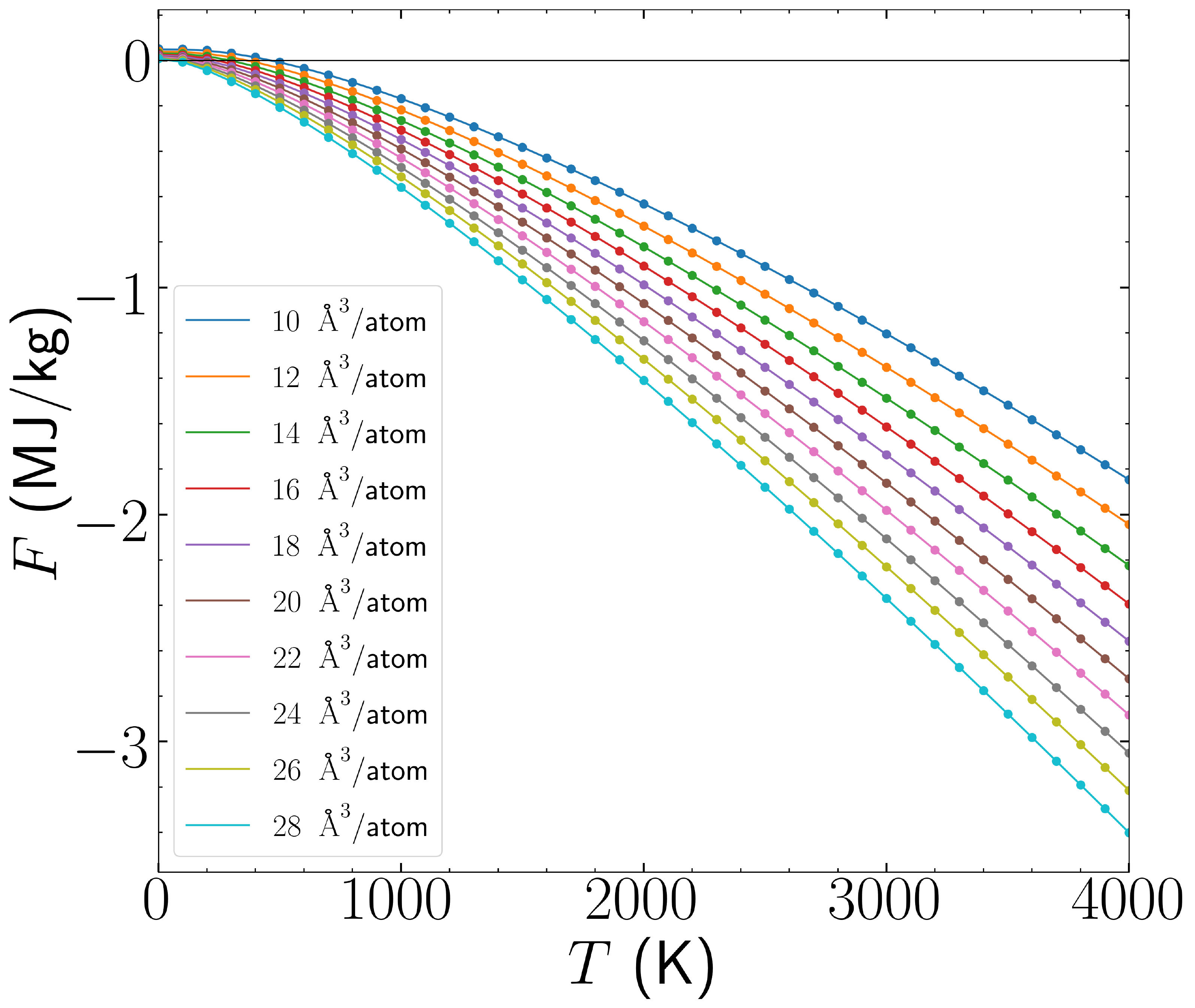}
  \caption{Free energy as a function of temperature for the $\gamma$
    phase computed from phonon calculations (dots) and the
    corresponding Debye model fit (lines) using Eq.~\ref{eq:min} over
    a range of different volumes.}
  \label{fig:3}
\end{figure}

We choose the Debye frequency $\nu_D$ via the minimization procedure
\begin{equation}
  \min_{\nu_D} \| F_D(T; V) - F_\mathrm{ph}(T;V)\|_2,
  \label{eq:min}
\end{equation}

where $F_D$ is the Debye free energy, $F_\mathrm{ph}$ is the free
energy from phonon calculations, $\|f(T)\|_2^2=\int_0^{T_\mathrm{max}}
f^2(T) dT$ denotes the $L_2$ norm, and we use $T_\mathrm{max}=4000$
K. The solution is obtained via a least-squares routine. Note that we
determine $\nu_D$ over a range of volumes, resulting in $\nu_D(V)$ for
each solid phase.  The determination of $\nu_D$ in this way leads to
very close agreement of $F_D$ and $F_\mathrm{ph}$ over a wide range of
volumes and temperatures, as shown in Fig.~\ref{fig:3}.  In
Fig.~\ref{fig:3}, the solid lines are computed using the Debye model,
while the dots are computed from phonon calculations.  We find close
agreement at low and high temperatures, with slightly poorer agreement
at low temperatures for small volumes. Also note that the $\nu_D$
determined using the minimization scheme above were in very close
agreement with the zeroth moment results in cases where the zeroth
moment results were not prone to numerical evaluation errors. The
advantage of the minimization approach is that it is not prone to the
same numerical difficulties.

The calculation of $\nu_D(V)$ for each phase provides a set of points
that can be used to fit an analytical form that corresponds to a
particular Gr\"{u}neisen model in OpenSesame. In this case, we use the
form
\begin{equation}
    \nu_D(V) = \nu_\mathrm{ref}\,x^{-A}\exp\left\{ -B (x-1) - {C\over 2}(x^2 - 1) \right\}
  \label{eq:chartdnu}
\end{equation}

where $x = V/V_\mathrm{ref}$ and $V_\mathrm{ref}$ is a reference
volume close to the equilibrium volume of each phase, to determine
parameters for the Gr\"{u}neisen model. Here the parameter $A$ is
chosen to be 2/3, while $B$ and $C$ are determined by fitting
Eq.~\ref{eq:chartdnu} to the Debye frequencies. The Gr\"{u}neisen
parameter $\Gamma = -\mathrm{d} \ln \nu_D/ \mathrm{d} \ln V$ is then
\begin{equation}
  \Gamma(V) = A + Bx + Cx^2.
  \label{eq:Gamma}
\end{equation}

With this form for $\Gamma$ and fixed $A$, there are effectively two
parameters that determine the Gr\"{u}neisen model for each phase, $B$
and $C$, and these are determined from DFT data by fitting
Eq.~\ref{eq:chartdnu} to DFT-calculated $\nu_D(V)$ points. Note,
however, that in OpenSesame, $B$ and $C$ are not used, but instead the
Gr\"{u}neisen parameter $\Gamma$ and its derivative $\Gamma'\equiv
\mathrm{d}\Gamma/\mathrm{d}\ln\rho$ are specified at a reference
volume $V_\mathrm{ref}$ (or equivalently, reference density
$\rho_\mathrm{ref}$).  Therefore, rather than specifying the
Gr\"{u}neisen model through $B$ and $C$, we use $\Gamma_\mathrm{ref}
\equiv \Gamma(\rho_\mathrm{ref})$ and $\Gamma_\mathrm{ref}' \equiv
\Gamma'(\rho_\mathrm{ref})$, with values listed in
Table~\ref{tab:2}. The Debye temperature $\Theta_D$ is also specified
at the same reference density $\rho_\mathrm{ref}$. For the 2162 EOS,
$\rho_\mathrm{ref}=7.581$ g/cm$^3$ is chosen for all phases, which is
within 1\% of the equilibrium density $\rho_0$ determined from the DFT
cold curve calculations. Also note that the form of $\Gamma$ in
Eq.~\ref{eq:Gamma} is only used for $V \leq V_\mathrm{ref}$. Above
$V_\mathrm{ref}$, another form for $\Gamma$ is used to prevent
$\Gamma$ from diverging in the expansion region well below solid
densities, such that $\Gamma \rightarrow 1$ as $V\rightarrow \infty$
(see Ref.~\onlinecite{rehn2020using} for additional details).  The
values of $\Theta_D(\rho_\mathrm{ref})$, $\Gamma_\mathrm{ref}$, and
$\Gamma'_\mathrm{ref}$ for each phase are shown in
Table~\ref{tab:2}. Values not in parentheses are the values computed
directly from DFT calculations, while values that were changed to fit
to experimental data (discussed in Sec.~\ref{sec:results}) are shown
in parentheses.

The advantage of phonon calculations is that they provide a way to
determine an initial set of Debye frequencies and Gr\"{u}neisen model
parameters in OpenSesame to generate the full tabular EOS. Additional
details on the process used to fit to DFT data are provided in
Ref.~\onlinecite{rehn2020using}.

\subsection{DFT-MD\label{sec:dftmd}}
All DFT-MD simulations were performed in VASP. We used an $NVT$
ensemble with a Nos\'e-Hoover
thermostat~\cite{nose1984unified,hoover1985canonical} to determine
liquid isotherms and an $NVE$ ensemble to determine the melt curve
using the $Z$
method.~\cite{belonoshko2006melting,burakovsky2014z,burakovsky2015ab}

For the $NVT$ simulations, we used the AM05 functional, a plane-wave
energy cutoff of 250 eV, a 2 fs time step, and 4 electrons in the
valence with the PAW method used to treat the core states. All
simulations used 250 atoms in a cubic cell (a $5\times5\times5$
supercell of the conventional bcc $\delta$ phase) and the Brillouin
zone was represented using the $\Gamma$ point only. A series of
simulations were run at different volumes and temperatures. Fermi
smearing was used for all simulations, setting the smearing width
according to the fixed temperature of each simulation. Each simulation
was started using the atomic positions of a liquid state that was
generated from a separate DFT-MD simulation in which the solid
$\delta$ phase was allowed to melt at 10,000 K over a period of time.
The resulting liquid structures were then used as the initial
structures at each $V,T$.  A total of 8 ps were used to determine the
resulting pressure, with the first 4 ps neglected to allow enough time
for the system to equilibrate. The pressure was then determined
through a time average of the last 4 ps using a block averaging
procedure (see the Supplemental Material for details). Additional
details on the DFT-MD calculations are provided in
Ref.~\onlinecite{rehn2020using}.

\begin{figure}
  \includegraphics[width=0.45\textwidth]{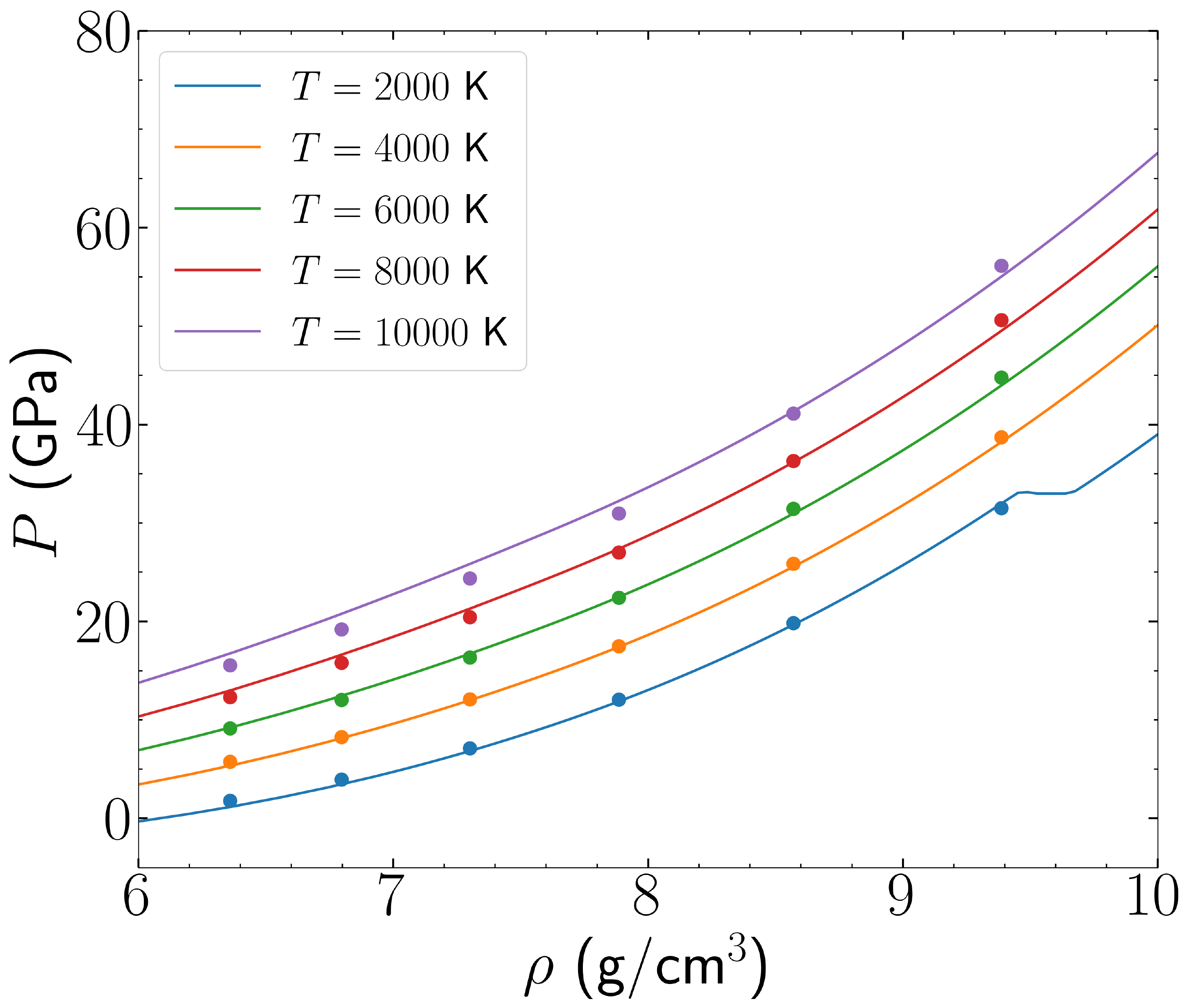}
  \caption{Isotherms from DFT-MD simulations of the liquid phase
    (dots) with a comparison to the 2162 EOS (lines).}
  \label{fig:4}
\end{figure}

Fig.~\ref{fig:4} shows liquid isotherms computed in DFT-MD (dots),
along with the isotherms used for the resulting 2162 EOS
(lines). Error bars are not included with the points because the
statistical errors in the pressure are small enough to fit within the
dots on the plot (see the Supplemental Material for details). The
isotherms help to constrain both the parameters controlling the
thermal response and parameters associated with the liquid in
OpenSesame. The resulting 2162 EOS fit is shown in the same figure,
with overall good agreement with the DFT-MD data. Some discrepancies
of the 2162 EOS and the DFT-calculated are present, which is a result
of having to determine liquid model parameters that simultaneously fit
the DFT-MD data and experimental Hugoniot data in the liquid phase, as
described in Sec.~\ref{sec:results}. We also point out that the
process for determining model parameters for the liquid phase is more
involved than that of the solid phases.  We leave this discussion for
Sec.~\ref{sec:results}, with additional details provided in
Ref.~\onlinecite{rehn2020using}.

For the $Z$-method simulations used to determine the melt curve, we
used an $NVE$ ensemble with the PBE functional and the PAW method in
VASP. Since the simulations were performed at high-$PT$ conditions, we
used accurate PAW potentials where the semi-core 4d states were
treated as valence states, so that each Sn atom included 14 valence
electrons per atom (4$d$, 5$s$, and 5$p$ orbitals). The valence states
were represented with a plane-wave energy cutoff of 300 eV.

\begin{figure*}
  \includegraphics[width=0.9\textwidth]{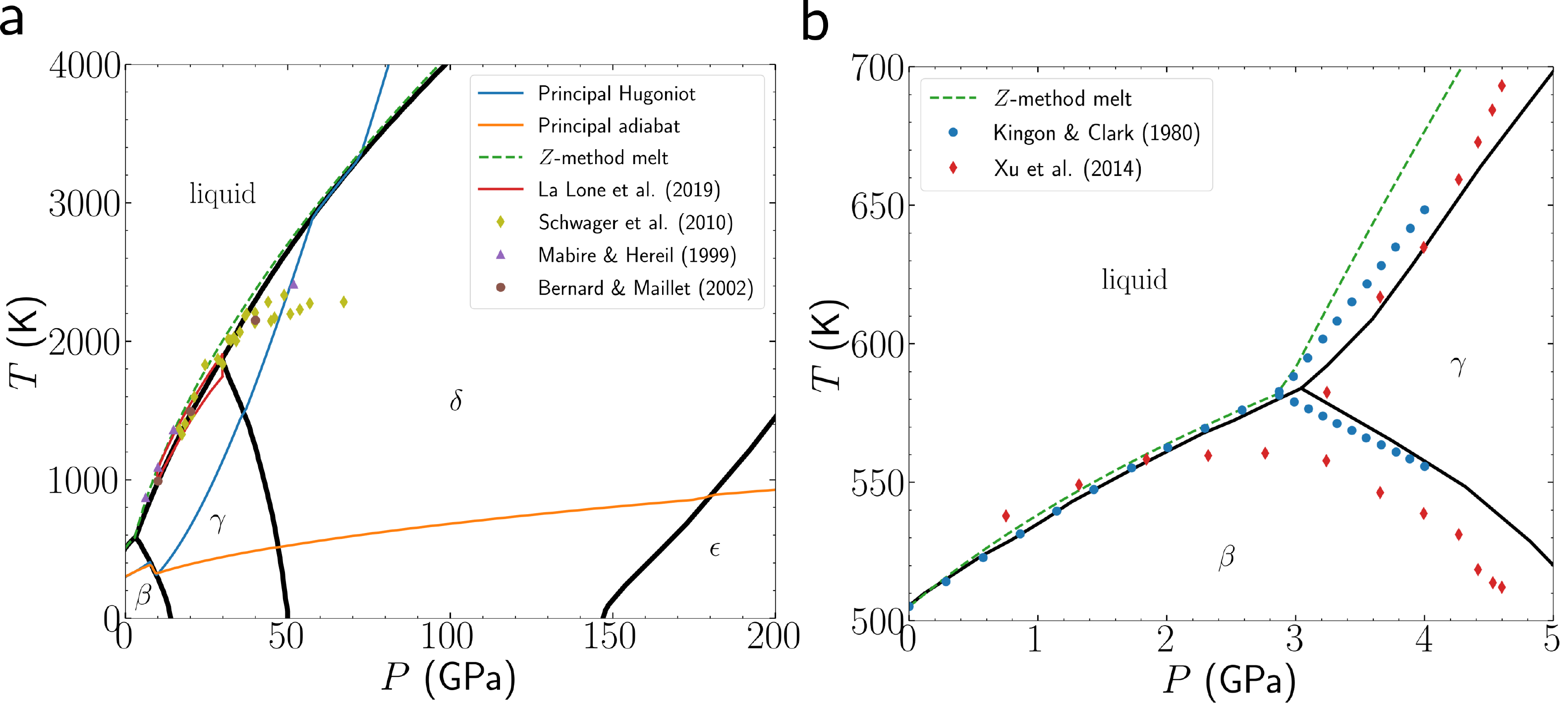}
  \caption{Phase diagram of tin. (a) A wide range of the state space
    with the principal Hugoniot (blue line) and principal adiabat
    (orange line). Experimental data of the melt curve are from
    Refs.~\onlinecite{la2019high,schwager2010melting,mabire2000shock,mabire2000shock2}.
    The data of Bernard \& Maillet are MD simulations discussed in
    Ref.~\onlinecite{bernard2002first}. (b) A narrow range of the
    state space around the $\beta$-$\gamma$-liquid triple
    point. Experimental data are from
    Refs.~\onlinecite{kingon1980redetermination,xu2014phase}.}
  \label{fig:5}
\end{figure*}

For all $Z$ method calculations involving non-cubic cells, we first
relaxed the structure to determine its unit cell parameters; those
unit cells were used for the construction of the corresponding
supercells. We used systems of sizes 512 ($4\times4\times8$) for the
$\beta$ phase, 504 ($6\times6\times7$) for the $\gamma$ phase, and 512
(rhombohedral $8\times8\times8$ with $\theta=109.5$ deg.) for the
$\delta$ phase. Only the $\Gamma$-point was used to represent the
Brillouin zone in each case. Full energy convergence (to $\sim1$
meV/atom) was verified by performing short runs with $2\times2\times2$
and $3\times3\times3$ k-point meshes and by comparing their output
with that of the run with a single $\Gamma$-point. The $Z$-method
$NVE$ runs were 15,000–20,000 timesteps, with a time step of 1
fs. These points were then used to determine an analytic form for the
melt curves of each solid phase, which are presented in the
Supplemental Material.  The full melt curve is the envelope of the
three melt curves for each individual phase and is shown in
Fig.~\ref{fig:5}.

\section{Results\label{sec:results}}
The DFT results presented in Sec.~\ref{sec:dft} provide a way to
determine an initial set of parameters for models used in OpenSesame.
This initial set of parameters provides a good starting point, but the
DFT data will not in general agree perfectly with experimental data.
The reasons for this have primarily to do with uncertainty in the
exchange-correlation functional used in DFT, as well as inherent
uncertainties in experiments.  Different experiments can give
significantly different results, making it necessary to prioritize
certain experimental results based on reasonable criteria when
constructing the EOS.  In this section, we briefly describe the
process used to adjust model parameters determined from DFT data to
fit experimental results, with a focus on the resulting fit to
experimental data. The final values before and after fitting are shown
in Table~\ref{tab:2}, where values \emph{not} in parentheses are
determined directly from DFT and values in parentheses indicate that
the value was adjusted to agree with experimental data. A more
detailed description of this process can be found in
Ref.~\onlinecite{rehn2020using}.

\subsection{Phase diagram}
We first review the overall phase diagram in $P$-$T$ space in the
final 2162 EOS, shown in Fig.~\ref{fig:5}.  The phase diagram
generated from DFT data alone does not fit perfectly with experimental
data, but the qualitative features are correct, as described in
Ref.~\onlinecite{rehn2020using}. The primary tool for adjusting the
location of the phase boundaries in pressure is by adjusting the cold
curve equilibrium energies $E_0=F_\mathrm{cold}(V_0)$. The
DFT-predicted values and adjusted values (in parentheses) are shown in
Table~\ref{tab:2}, where the energy is relative to $E_0$ of the
$\beta$ phase. Although the changes appear to be relatively large,
there are two important reasons for this. First, it is not possible
with GGA xc functionals to simultaneously predict correct atomization
energies and bond lengths;~\cite{perdew2008restoring} the prediction
of both is only possible within a meta-GGA framework. The AM05
functional used for DFT calculations in Sec.~\ref{sec:dft} gives very
accurate predictions of bond lengths for tin, but because it is a GGA,
should not be expected to provide accurate relative values of
$E_0$. Second, adjusting $E_0$ does not change any properties that
depend on derivatives of the energy, making $E_0$ the most
straight-forward and least problematic parameter to use to adjust the
relative locations of the phase boundaries in $P$-$T$ space.

Several points regarding the phase diagram should be noted. First, the
$\beta$-$\gamma$ phase boundary was modified from the SESAME 2161 EOS
to be located at slightly lower pressures.  The 2161 EOS originally
determined the phase boundary based on shock Hugoniot data, which
included some kinetic effects associated with hysteresis in the
$\beta\rightarrow\gamma$ phase transition. With recent developments in
modeling the kinetics of phase transitions in hydrodynamics codes, it
is now important to place the phase boundary at the location of static
experiments so that the kinetic effects associated with the
$\beta$-$\gamma$ phase transition can be accounted for in the kinetic
models, rather than in the EOS. This allows the reverse transformation
($\gamma\rightarrow\beta$) to properly include kinetic effects, as
well.  The $\beta$-$\gamma$ phase boundary and the
$\beta$-$\gamma$-liquid triple point are shown in Fig.~\ref{fig:5}b
with comparisons to experimental data from
Refs.~\onlinecite{kingon1980redetermination,xu2014phase}. The 2162 EOS
is in close agreement with the experimental data near the triple point
and with the $\beta$-$\gamma$ phase boundary. The higher pressure
parts of the phase boundary are also in close agreement with phase
diagrams reported previously in
Refs.~\onlinecite{tonkov2018phase,young1991phase,cox2015fitting}.

We also point out that the slope of the $\gamma$-$\delta$ phase
boundary in Fig.~\ref{fig:5}a is determined entirely from DFT
data. One challenge in determining this phase boundary lies in the
fact that the $\gamma$ phase cold curves should in principle be
computed in such a way that the lattice is allowed to relax at each
volume. This leads to a $c/a$ ratio that converges to 1 as the volume
is decreased, indicating that the $\delta$ (bcc) phase becomes lower
in energy.  Although in principle this is the correct way to treat the
$\gamma$ phase, in practice this makes it very difficult to determine
a well-defined $\gamma$-$\delta$ phase boundary. At elevated
temperatures the Gibbs free energy $G(P)$ of the $\gamma$ and $\delta$
phases are nearly coincident near the phase boundary, making it
difficult to numerically determine the pressure $P$ associated with
the transition at elevated temperatures.  The ambiguity in defining a
phase boundary is also borne out in experiments. In particular,
Salamat \emph{et al.}  find in Ref.~\onlinecite{salamat2013high}
evidence of the emergence of body-centered orthorhombic (bco) phase(s)
near the $\gamma$-$\delta$ phase transition at room temperature. This
finding indicates that near the phase boundary, the relative energy
differences between bct, bco, and bcc structures are small enough that
the presence of local strains within different grains of the sample
can lead to the local stabilization of different phases. These results
provide insight into the fundamental difficulty of defining a clear
phase boundary in this region of the state space, rather than indicate
a fundamental failure in the DFT calculations.

In order to address the difficulties associated with defining the
$\gamma$-$\delta$ phase boundary, we ultimately used a DFT cold curve
for the $\gamma$ phase in which the $c/a$ ratio is kept fixed at the
equilibrium (zero pressure) value.  This allows for more contrast in
the enthalpies (at $T=0$) and the Gibbs free energies ($T>0$) near the
phase boundary and therefore alleviates difficulties associated with
the numerically ambiguous phase preference near the phase boundary.

Another point regarding the $\epsilon$ phase is also worth
mentioning. The experimental evidence for the emergence of the
$\epsilon$ phase is relatively limited, but DAC experiments by Salamat
\emph{et al.} (Ref.~\onlinecite{salamat2011dense}) show a clear
transition to the $\epsilon$ phase at room temperature above 150 GPa.
Nonetheless, these experiments do not provide information about the
slope of the phase boundary in $P$-$T$ at these high pressures, making
the DFT-predicted phase boundary the best currently available data we
have. The DFT results clearly indicate a positive slope in
Fig.~\ref{fig:5}a which is also in qualitative agreement with
experiments performed by Lazicki \emph{et al.}
(Ref.~\onlinecite{lazicki2015x}) involving dynamic compression of tin
up to pressures around 1 TPa. In Ref.~\onlinecite{lazicki2015x}, no
evidence of the $\epsilon$ phase is found, however it is noted that
very high temperatures are achieved during compression, which may mean
that the $(P,T)$ states probed in experiments are located above the
$\delta$-$\epsilon$ phase boundary that we compute using
DFT. Furthermore, Lazicki \emph{et al.} postulate that strain rates in
their compressive experiments may be too fast to observe nucleation
and growth of the $\epsilon$ phase.  New compressive experiments would
help to shed light on these issues and on the location of the phase
boundary.

\begin{figure*}
  \includegraphics[width=0.95\textwidth]{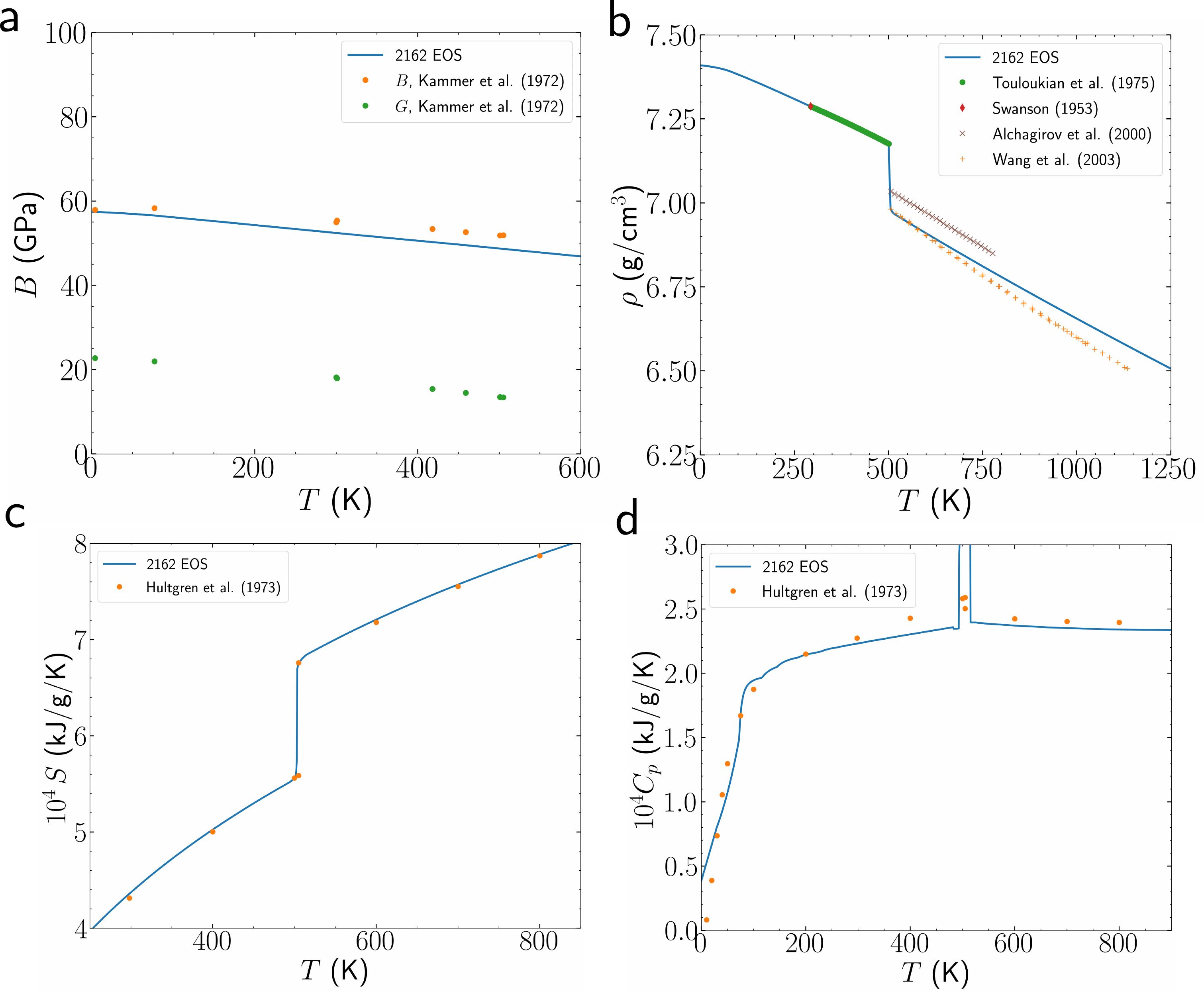}
  \caption{Comparison of the 2162 EOS to experimental isobaric data
    along the 1 bar isobar. (a) The bulk modulus $B(T)$ and shear
    modulus $G(T)$ are computed from temperature-dependent elastic
    constant data in Ref.~\onlinecite{kammer1972elastic}. (b) The
    density $\rho(T)$ in the $\beta$ phase (below 505 K) and in the
    liquid phase (above 505 K) are compared to experimental data from
    Refs.~\onlinecite{touloukian1975thermal,swanson1953standard,alchagirov2000temperature,wang2003precise}. Experimental
    data from Ref.~\onlinecite{hultgren1973selected} are compared with
    the 2162 EOS for (c) the entropy $S(T)$ and (d) the heat capacity
    at constant pressure $C_p(T)$.}
  \label{fig:6}
\end{figure*}

One final point regarding the melt curve is important. Many
measurements of the melt curve have been made using a variety of
experimental techniques, with a large amount of scatter in the
data. Due to the degree of variability, it is necessary to choose a
subset of experimental results to guide the EOS construction.  For the
2162 EOS we have focused on recent experiments by La Lone \emph{et
  al.}  (Ref.~\onlinecite{la2019high}) which measure melt on shock
release using a combination of pyrometry, reflectance, and velocimetry
techniques to determine the location of the melt curve in $(P,T)$
space. These results are shown in Fig.~\ref{fig:5}a, where the red
lines indicate upper and lower limits of uncertainty in the
measurements. We also include laser-heated DAC data by Schwager
\emph{et al.}  (Ref.~\onlinecite{schwager2010melting}), shock-induced
melt data by Mabire \& H\'{e}reil
(Refs.~\onlinecite{mabire2000shock,mabire2000shock2}), and classical
molecular dynamics (MD) simulations of the melt by Bernard \& Maillet
(Ref.~\onlinecite{bernard2002first}). Note that the calculations of
Bernard \& Maillet use an interatomic potential optimized by fitting
to DFT-MD simulations. We also show the melt curve predicted by our
$Z$ method calculations in Fig.~\ref{fig:5}, which are in very close
agreement to the 2162 EOS. Comparisons to other experiments can be
found in Ref.~\onlinecite{rehn2020using}.  It is also important to
point out that the melt curve of the 2162 EOS is located at slightly
higher temperatures than the previous 2161 EOS, based primarily on the
experimental data provided in Fig.~\ref{fig:5} and also on a trend
towards consensus that the melt curve lies at higher temperatures than
what is shown in the 2161 EOS.~\cite{la2019high,cox2015fitting}

The melt curve is determined primarily by the liquid phase parameters,
shown in Table~\ref{tab:2}, in addition to values of liquid-specific
parameters $\Gamma_{\rm m}, \Gamma_{\rm rs}, T_{\rm m}, T_{\rm rs}$,
and $\Delta S$, described in Sec.~\ref{sec:opensesame}. For these
liquid-specific parameters, we use the values $\Gamma_{\rm rs} =
\Gamma_{\rm m} = 2.45$, which is the same as the adjusted
$\Gamma_\mathrm{ref}$ value for the $\gamma$ phase, as well as $T_{\rm
  m} = T_{\rm rs} = 580$ K and $\Delta S = 0.9$ kJ/g.  Note that all
parameters for the liquid phase in Table~\ref{tab:2} are shown in
parentheses because there is not a unique way to determine the
parameters from DFT calculations.  The liquid phase parameters are the
least straight-forward to adjust, and we provide a brief overview of
how these parameters were determined below (additional details are
provided in Ref.~\onlinecite{rehn2020using}). First, because the melt
curve extends across the $\beta$, $\gamma$, and $\delta$ phases, it
was necessary to construct a cold curve that interpolated between
these three phases. This was done by stitching together $P(\rho)$ for
the $\beta$ and $\gamma$ phases near the $T=0$ transition density (the
$\delta$ phase is left out because the $\gamma$ and $\delta$ $P(\rho)$
curves are nearly overlapping at high density), so that an `effective'
cold curve that interpolates between phases could be used. The
resulting cold curve has a slightly lower $\rho_0$ than the $\beta$,
$\gamma$, and $\delta$ phases, as shown in Table~\ref{tab:2}. The
values of $B$ and $B'$ are also determined from fitting the Vinet-Rose
model in this way.  The parameters $E_0$ and $\Theta_D$ for the liquid
phase mainly influence the location of the melt curve in $P$-$T$
space.  Because there is no straight-forward way to determine these
parameters directly from DFT data, the adjustment was done mostly
through trial-and-error fitting to the experimental data of melt curve
measurements. Lastly, the Gr\"{u}neisen parameter
$\Gamma_\mathrm{ref}$ was initially chosen to be the same as
$\Gamma_\mathrm{ref}$ of the $\gamma$ phase. However, this resulted in
pressures that were slightly too high at elevated temperatures in both
the DFT-MD data (Fig.~\ref{fig:3}) and in the shock Hugoniot data
(Fig.~\ref{fig:8}). Lowering this value by $\sim2$\% allowed for
better agreement in both the DFT-MD results and the shock Hugoniot
results. As described in Sec.~\ref{sec:shock}, lowering this value
changed the location of the phase boundaries in Fig.~\ref{fig:5}
slightly. We therefore reduced $\Gamma_\mathrm{ref}$ of the solid
phases by the same ratio as was used for the liquid phase to restore
the accuracy of the phase boundaries.  Note that
$\Gamma'_\mathrm{ref}$ of the liquid phase primarily controls the
curvature of the melt curve at high pressures. We determined the value
$\Gamma'_\mathrm{ref} =-3.5$ in Table~\ref{tab:2} by fitting to the
melt curve at high pressure. Because this value was relatively higher
than that of the solid phases, $\Gamma'_\mathrm{ref}$ of the solid
phases were decreased to retain the shapes of the phase boundaries at
high pressures.

\subsection{Isobaric data}
When adjusting model parameters in OpenSesame to create the 2162 EOS,
we started with isobaric data. As described in
Ref.~\onlinecite{rehn2020using}, the isobaric data generally allows
for some constraints to be placed on model parameters of the phases
that are stable at atmospheric pressure. For tin, this includes the
$\beta$ and liquid phases.

\begin{figure*}
  \includegraphics[width=\textwidth]{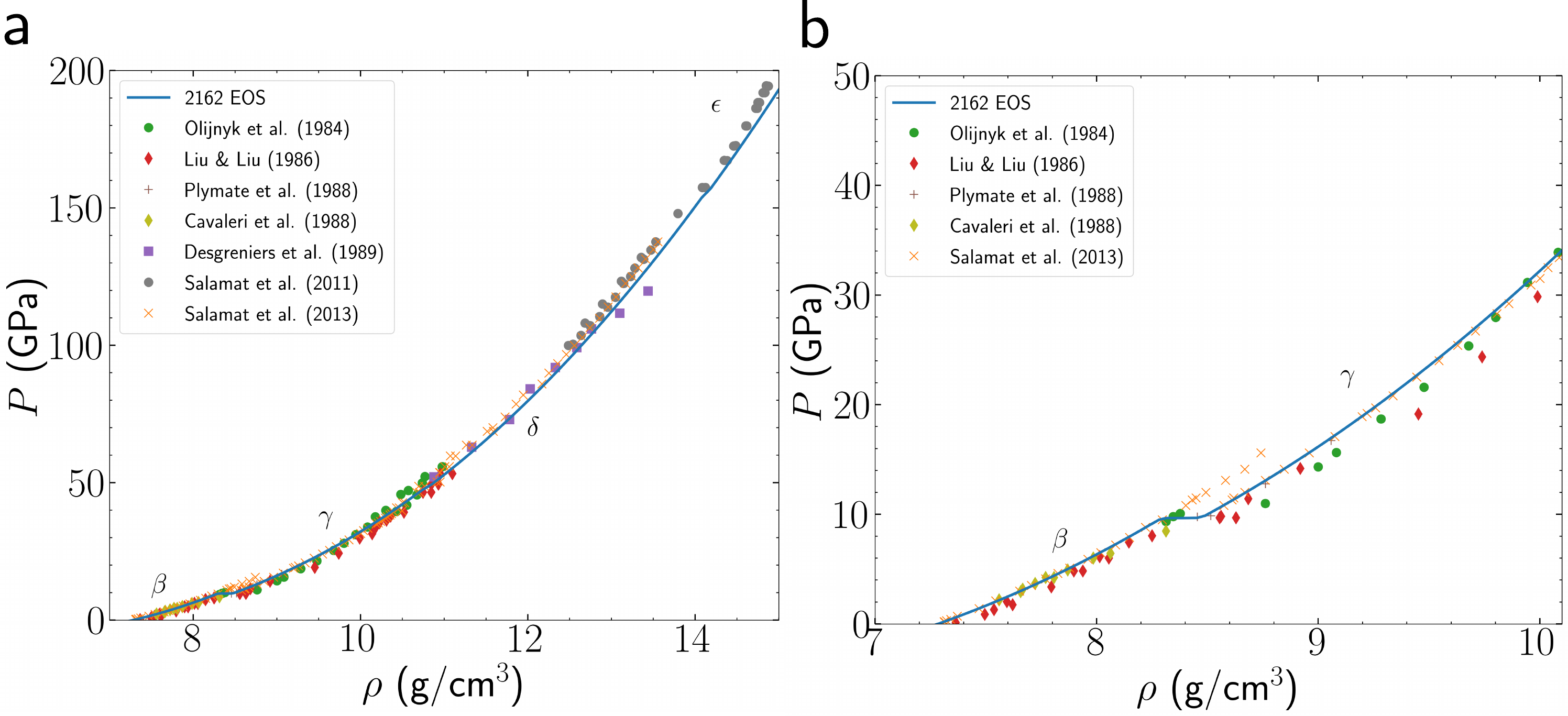}
  \caption{Room temperature isotherm for the 2162 EOS with a
    comparison to experimental data over (a) the full range of
    experimental data and (b) near the $\beta$-$\gamma$ phase
    transition. Experimental data are from
    Refs.~\onlinecite{olijnyk1984phase,liu1986compressions,plymate1988pressure,cavaleri1988pressure,desgreniers1989tin,salamat2011dense,salamat2013high}.}
  \label{fig:7}
\end{figure*}

In Fig.~\ref{fig:6}, we show a variety of isobaric data at 1 bar and
at temperatures above and below the $\beta$-liquid phase boundary. In
Fig.~\ref{fig:6}a the temperature dependent bulk modulus $B(T)$ of the
$\beta$ phase is shown with comparisons to experimental data
determined from Kammer \emph{et al.}
(Ref.~\onlinecite{kammer1972elastic}). Kammer \emph{et al.} measured
the elastic constants of $\beta$ tin at different temperatures, and
from those measurements it is possible to estimate the bulk modulus
$B$ and shear modulus $G$, as described in
Ref.~\onlinecite{rehn2020using}. Using the data of Kammer \emph{et
  al.}, we found that at $T=0$, $B=58$ GPa, which can be compared to
the AM05 result of $B=52.7$ GPa, as found by fitting the $E$-$V$
points to a Vinet-Rose model.~\cite{vinet1987temperature} Because the
DFT result is slightly low, we modified the value to $B=58$ GPa (see
Table~\ref{tab:2}), which provides a constraint that reduces the
number of remaining model parameters that need to be adjusted.

In Fig.~\ref{fig:6}b, we show the density along the 1 bar isobar with
comparisons to experimental data from
Refs.~\onlinecite{touloukian1975thermal,swanson1953standard,alchagirov2000temperature,wang2003precise}.
The 2162 EOS matches very well in the $\beta$ phase below melt, and
lies in between the data of Alchagirov \emph{et al.} and Wang \emph{et
  al.} in the liquid phase.  Due to the constraint on $B$ at $T=0$, we
were able to determine the equilibrium density of the $\beta$ phase,
$\rho_0=7.437$ g/cm$^3$ based on the $\rho(T)$ data in
Fig.~\ref{fig:6}b. This value is roughly 1.2\% higher than the
AM05-predicted value of 7.344 g/cm$^3$ (see Table~\ref{tab:2}) and
7.2\% higher than the PBE-predicted value of 6.939 g/cm$^3$. The other
piece that we adjusted slightly to fit to the isobaric $\rho(T)$ data
is $\Gamma_\mathrm{ref}$, shown in Table~\ref{tab:2}. The combination
of $\rho_0$, $B$, and $\Gamma_\mathrm{ref}$ together control the slope
and initial value of $\rho(T)$, and these values were found to give
good agreement with data shown in Fig.~\ref{fig:6}b.

We also show the entropy (Fig.~\ref{fig:6}c) and heat capacity at
constant pressure (Fig.~\ref{fig:6}d) with comparisons to data from
Hultgren \emph{et al.} (Ref.~\onlinecite{hultgren1973selected}). Very
close agreement is shown between the experimental data and the 2162
EOS. No additional adjustment of model parameters was necessary to
provide good agreement for the $\beta$ phase.

\subsection{Isothermal data}
The next step in the EOS construction is to look at isothermal DAC
data at room temperature. A variety of DAC experiments have been
performed over the years, and we focus specifically on data from
Refs.~\onlinecite{olijnyk1984phase,liu1986compressions,plymate1988pressure,cavaleri1988pressure,desgreniers1989tin,salamat2011dense,salamat2013high}.
In Fig.~\ref{fig:7} we show a comparison of the 2162 EOS to these
experimental results. Note that due to the high level of agreement of
the AM05 cold curves with DAC data (Fig.~\ref{fig:1}), it was possible
to use the AM05 cold curves directly and without any adjustments,
aside from the $\beta$ phase which was modified slightly based on
isobaric data. The unmodified cold curve parameters are shown in
Table~\ref{tab:2}.

In Fig.~\ref{fig:7}a, we show results over a wide range of density,
which spans all four solid phases included in the EOS, with the
$\beta$-$\gamma$ transition occurring at roughly 9 GPa, the
$\gamma$-$\delta$ transition occurring at roughly 48 GPa, and the
$\delta$-$\epsilon$ transition occurring at roughly 156 GPa. The
experiments of Salamat \emph{et al.}
(Refs.~\onlinecite{salamat2011dense,salamat2013high}) are more recent
and are the most precise data available, and the 2162 EOS shows a high
level of agreement with these results, particularly in the $\beta$ and
$\gamma$ phases at low pressure, as shown in Fig.~\ref{fig:7}b.  Note
that small discrepancies can be seen in the $\delta$ phase in
Fig.~\ref{fig:7}a, where the 2162 EOS is slightly lower in pressure
than the results of Salamat \emph{et al.}. This is interesting due to
the fact that the discrepancies start to appear at the onset of the
$\gamma$-$\delta$ phase transition. In the experiments of
Ref.~\onlinecite{salamat2013high}, Salamat \emph{et al.} find evidence
for the emergence of bco phases, indicating that the $\gamma$ (bct),
$\delta$ (bcc), and bco phases are all accessible in this region of
the state space and that the observed stable phase may be influenced
by small local strains present in the samples. For the purposes of the
EOS, we did not try to address all of these issues, but rather relied
on the AM05-calculated values, since this provided a concrete way for
determining model parameters without leading to significant
discrepancies with the available experimental data.
\begin{figure*}
  \includegraphics[width=\textwidth]{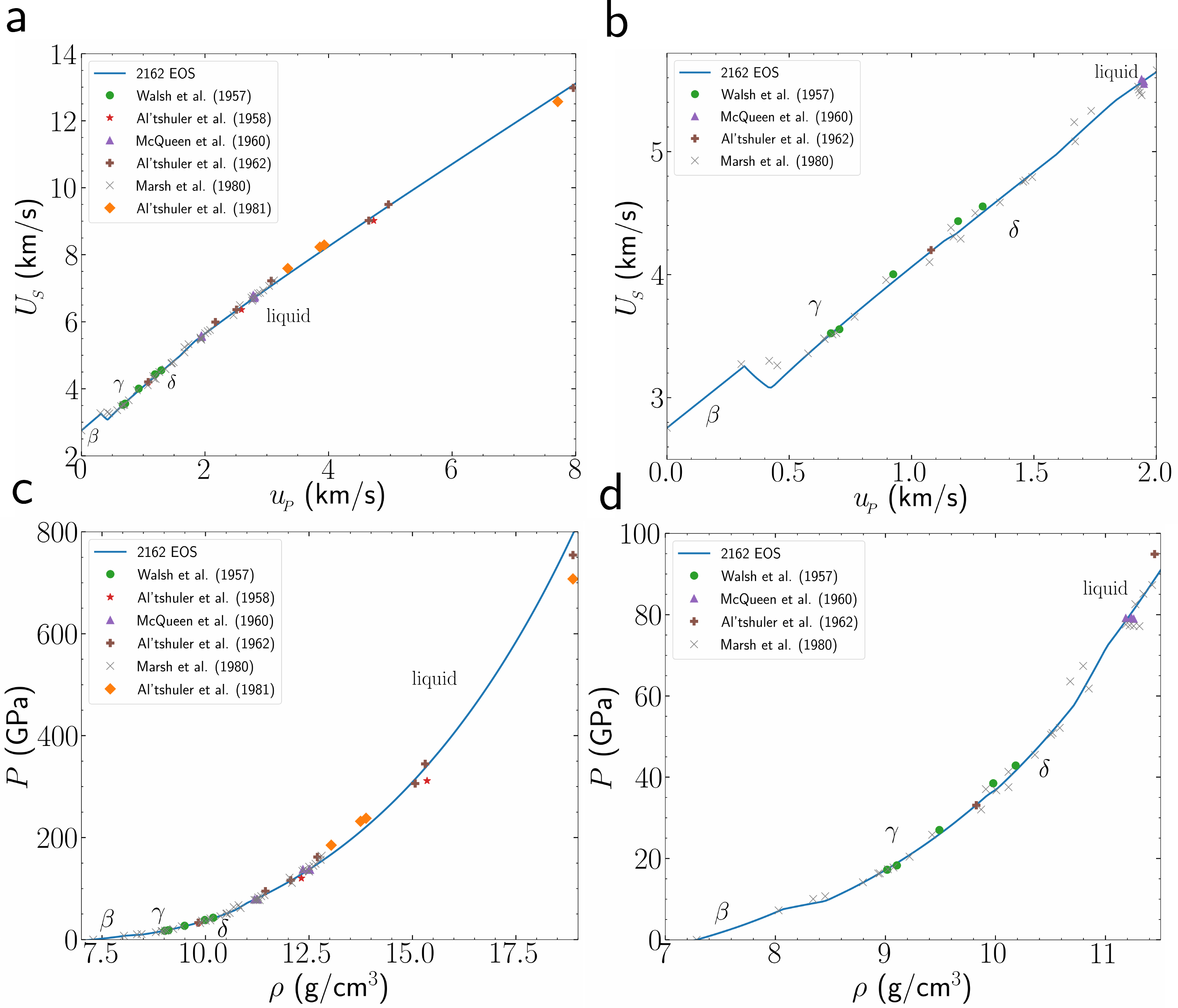}
  \caption{Comparison of the 2162 EOS to shock experiments along the
    principal Hugoniot. The $U_s$-$u_p$ relations are shown over (a) a
    wide range of shock speeds including the liquid phase and (b) over
    a narrow range of shock speeds focusing on the solid phases. The
    pressure $P(\rho)$ along the Hugoniot is shown over (c) a wide
    range of pressure and (d) a narrow range of pressure focusing on
    the solid phases. Experimental data are from
    Refs.~\onlinecite{walsh1957shock,al1958dynamic,mcqueen1960equation,al1962shock,marsh1980lasl,al1981shock}.}
  \label{fig:8}
\end{figure*}

\subsection{Shock data\label{sec:shock}}
The final piece of experimental data we consider is shock
data. Fig.~\ref{fig:8} shows a comparison of the 2162 EOS to
experimental shock measurements from
Refs.~\onlinecite{walsh1957shock,al1958dynamic,mcqueen1960equation,al1962shock,marsh1980lasl,al1981shock}. Note
that the principal Hugoniot is shown in Fig.~\ref{fig:5}a. In
Figs.~\ref{fig:8}a-b we show the $U_{\!_S}$-$u_{\!_P}$ relations along
the principal Hugoniot, where panel (a) shows a wide range of shock
speeds into the liquid phase and panel (b) is zoomed in to focus on
the solid phases. The 2162 EOS is in good agreement throughout. Note
that in panel (b), the $\beta\rightarrow\gamma$ phase transition is
shown to initiate at lower shock speed than the data points of the
Marsh et al. (1980) results.  As mentioned previously, this is due to
the fact that some hysteresis associated with the phase transition is
included in the shock data.  The goal of the 2162 EOS is to place the
phase boundaries at their equilibrium values and allow for
hydrodynamics codes to capture the kinetic effects associated with
phase transitions.  We also point out that the $\gamma$-$\delta$ phase
transition is very subtle, but can be seen at $U_{\!_S}\approx4.2$
km/s in panel (b). This is not expected to be a large effect since the
$\gamma$ phase exhibits a shear deformation to the $\delta$ phase
along the phase boundary. One interesting remaining question is if
there is any noticeable change in material strength through this phase
transformation.

In Figs.~\ref{fig:8}c-d, we show the corresponding pressure dependence
$P(\rho)$ along the Hugoniot. The 2162 EOS is again in good agreement
with the experimental data through the three solid phases and into the
liquid.

The results of Fig.~\ref{fig:8} did not require any additional
adjustments to the solid phases, but did require adjustments to the
Gr\"{u}neisen parameters of the liquid phase.  Originally, the liquid
phase was constructed using the same $\Gamma_\mathrm{ref}$ and
$\Gamma'_\mathrm{ref}$ values as the $\Gamma$ phase. However, we found
that this value of $\Gamma_\mathrm{ref}$ caused $U_{\!_S}$ to be
slightly too high at large $u_{\!_P}$ in Fig.~\ref{fig:8}a and also
caused the pressure to be slightly too high at high densities in
Fig.~\ref{fig:8}c.  In order to address these issues, we lowered
$\Gamma_\mathrm{ref}$ for the liquid phase from 2.48 to 2.45 (see
Table~\ref{tab:2}), a change of roughly 2\%, which brought the
Hugoniot down in both $U_{\!_P}$ and $P$ and resulting in the
agreement with experimental data shown in Fig.~\ref{fig:8}. However,
because of these adjustments, the solid-solid and solid-liquid phase
boundaries in Fig.~\ref{fig:5} also changed to be in worse agreement
with experimental data.  To address this, we ended up decreasing the
$\Gamma_\mathrm{ref}$ values of the $\gamma$, $\delta$, and $\epsilon$
phases by the same ratio as was used for the liquid phase (see
Table~\ref{tab:2}), which restored the agreement of the phase
boundaries with experimental data.

\section{Conclusion\label{sec:conclusion}}
We have described the construction of a new multiphase SESAME EOS for
tin, referred to as SESAME 2162. The new EOS includes four solid
phases and the liquid phase. We performed DFT calculations using the
AM05 exchange-correlation functional of the four solid phases and the
liquid phase, including cold curve and quasi-harmonic phonon
calculations of the solid phases and DFT-MD calculations of the liquid
phase.  The DFT calculations greatly aid in constraining the model
parameters used in OpenSesame to generate the resulting 2162
EOS. Because model parameters determined from DFT alone are not in
exact agreement with experimental data, slight adjustments of model
parameters determined by the DFT calculations are required.  We
presented the results of the 2162 EOS construction with comparisons to
a wide range of experimental data, including isobaric data, isothermal
data, shock data, measurements of the triple point and solid-solid
phase boundaries, and measurements of the melt curve.  The 2162 EOS
shows an overall high level of agreement with these experimental
results.  In addition, in regions of the state space where
experimental data is limited or does not exist, the DFT calculations
provide the best available information on material properties. Looking
forward, it will be important to develop new methods for EOS
construction, such as automated EOS generation based on both DFT and
experimental data. At the same time, uncertainty quantification of EOS
will be important. We expect that the process used to generate the
2162 EOS can be used to inform on both uncertainty quantification and
methods for automatic EOS generation, as described in more detail in
Ref.~\onlinecite{rehn2020using}.

\section*{Acknowledgements}
We would like to thank Ann Mattsson for useful discussions regarding
DFT calculations.  We also thank Sven Rudin for useful discussions
regarding phonon calculations.

This work was supported by Advanced Simulation and Computing, Physics
and Engineering Models, at Los Alamos National Laboratory.  Los Alamos
National Laboratory, an affirmative action/equal opportunity employer,
is managed by Triad National Security, LLC, for the National Nuclear
Security Administration of the U.S. Department of Energy under
contract 89233218CNA000001.

\bibliography{refs}{}

\begin{thebibliography}{120}
\bibitem [{\citenamefont {Burakovsky}()}]{Leonid-private-communication}%
  \BibitemOpen
  \bibfield  {author} {\bibinfo {author} {\bibfnamefont {Leonid}\ \bibnamefont
  {Burakovsky}},\ }\href@noop {} {}\bibinfo {howpublished} {To be
  published.}\BibitemShut {Stop}%
\end{thebibliography}

\end{document}